\documentclass[aps,prc,twocolumn,floatfix,showpacs,preprintnumbers,amsmath,amssymb,nofootinbib]{revtex4-1}
\usepackage[T1]{fontenc}
\usepackage{graphicx}
\usepackage{bm}
\usepackage{color}
\usepackage{dcolumn}
\usepackage{amssymb}
\usepackage[outdir=./]{epstopdf}

\newcommand{\fm}{{\rm\,fm}}
\newcommand{\MeV}{{\rm\,MeV}}
\newcommand{\fmd}{{\rm\,fm^{-3}}}
\newcommand{\fmk}{{\rm\,fm^{-1}}}
\newcommand{\I}{\mathrm{i}}
\renewcommand{\vec}[1]{\mbox{\boldmath $#1$}}
\newcommand{\vecs}[1]{\mbox{\boldmath {\scriptsize $#1$}}}
\begin{document}

\title{Twist-averaged boundary conditions for nuclear pasta Hartree-Fock calculations}

\author{B.~Schuetrumpf}
\affiliation{NSCL/FRIB Laboratory, Michigan State University, East
  Lansing, MI 48824, USA} 
\author{W.~Nazarewicz} 
\affiliation{Department of Physics and
  Astronomy and NSCL/FRIB Laboratory, Michigan State University, East
  Lansing, MI 48824, USA} 
  \affiliation{Faculty
  of Physics, University of Warsaw, Pasteura 5, 02-093 Warsaw,
  Poland}
  
\date{\today}

\begin{abstract}
\textbf{Background:} Nuclear pasta phases, present in the inner crust of neutron stars,
are associated with  nucleonic matter at sub-saturation densities arranged in regular shapes. Those complex phases, residing in a  layer which is approximately 100 m thick, impact many  features of neutron stars.
Theoretical quantum-mechanical simulations of nuclear pasta are usually carried out in finite 3D boxes assuming periodic boundary conditions (PBC). The resulting solutions are affected by spurious finite-size effects. 

\textbf{Purpose:} In order to remove  spurious finite-size effects, it is convenient to employ
 twist-averaged  boundary conditions (TABC) used in condensed matter, nuclear matter, and lattice QCD applications. In this work, we study the effectiveness of TABC in the context of pasta phases simulations within nuclear density functional theory.

\textbf{Methods:} We perform Skyrme-Hartree-Fock calculations in three dimensions  by implementing  Bloch boundary conditions. The  TABC averages are obtained by means of Gauss-Legendre integration over  twist angles. 

\textbf{Results:} We benchmark the TABC for a free nucleonic gas and apply it to simple cases such as the  rod and slab phases, as well as to more elaborate P-surface and gyroidal phases. 

\textbf{Conclusions:} We demonstrate that by applying TABC reliable results can be obtained from calculations performed in relatively small volumes. By studying various contributions to the total energy, we gain insights into pasta phases in mid-density range. 
\end{abstract}


\maketitle

\section{Introduction}

Nuclear matter as present on earth in the center of atoms is almost isotropic with a central density of $\rho_{\rm sat}\approx 0.16\fmd$, the nuclear saturation density. This changes drastically in astrophysical environments such as neutron stars or core-collapse supernova. In particular,  in the inner crust of neutron stars at sub-saturation densities  $0.1\rho_{\rm sat}< \rho < 0.8\rho_{\rm sat}$, nucleonic matter is expected to form complex structures commonly referred to as ``pasta'' phases \cite{Ravenhall,Hashimoto}. Because of low proton fractions and macroscopic dimensions,  pasta phases represent  a unique environment, which is not present on earth and cannot be recreated in the laboratory.

The occurrence and topology of pasta structures can have multiple influences on the properties of a neutron star and the evolution of a supernova. Not only it affects the neutrino transport \cite{Horowitz2004,Horowitz20042,Sonoda2007,Gry10}, but also the neutron star cooling \cite{Gusakov} and r-mode instabilities in rotating neutron stars \cite{Bildsten,Lindblom00}.

Nuclear pasta simulations can be divided into two categories.
The first group represents semi-classical methods; it includes approaches such as the
classical molecular dynamics \cite{Dorso2012}, Thomas-Fermi approach \cite{Williams1985,Oka13a,Pais15}, and quantum molecular dynamics (QMD) \cite{Schneider13,Schneider14,Maruyama,Sonoda2008,Watanabe09}.
The second family includes quantum-mechanical simulations based on nuclear density functional theory \cite{Bender03}, such as  Hartree-Fock (HF) calculations (with or without BCS pairing) \cite{Bonche,Mag02,Goegelein,NewtonStone,Pais12, Schuetrumpf2013a,Schuetrumpf2014,Schuetrumpf2015}. 
While current advanced QMD calculations can be performed with hundreds of  thousands of nucleons in  3D boxes greater than 100\,fm in size, self-consistent HF calculations  are still limited to 
$\sim$2,000 fermions in box-sizes of the order of 20\,fm. 

Since the  large-scale QMD simulations yield periodic pasta phases, it is reasonable to assume periodic boundary conditions in HF calculations. In most applications,  the HF wave functions are also constrained  to be periodic. This can lead to severe restrictions: Firstly, the solution has no chance to develop disorder on a  scale larger than the box length  \cite{Horowitz2015}.
This can only be remedied by using much larger volumes. 
Secondly, the periodic ansatz for the wave functions is fairly restrictive, because the most general solutions for a periodic potential are Bloch waves, which differ by a phase when moving to a neighboring box. If strictly periodic wave functions are considered, spurious finite-size (or shell) effects appear due  to the quantization of waves in the box. A method to remove the spurious finite-volume corrections is to employ twist-averaged boundary conditions, where  the observables are obtained by averaging over different Bloch phases (or twist angles) \cite{Gros1992,Gros1996,Lin}. A method based on TABC (sometimes referred to as  `integration over boundary conditions') has  been applied  to circumvent the need for large-volume boxes in the context of many-electron systems \cite{Gros1992,Gros1996,Lin}, nucleonic matter  \cite{Hagen14}, and lattice QCD \cite{Tiburzi05,Briceno14}. It was also used to describe the crust in neutron stars within mean-field models  \cite{Carter,Chamel2007} and  Quantum Monte Carlo approach \cite{Gulminelli}. The aim of this work is to apply  TABC to self-consistent HF pasta calculations to assess the final-volume effect on previous results obtained with PBC \cite{Schuetrumpf2013a,Schuetrumpf2014,Schuetrumpf2015} and estimate the realistic box sizes with TABC for future applications. 

\begin{figure}[tb]
\includegraphics[width=0.9\linewidth]{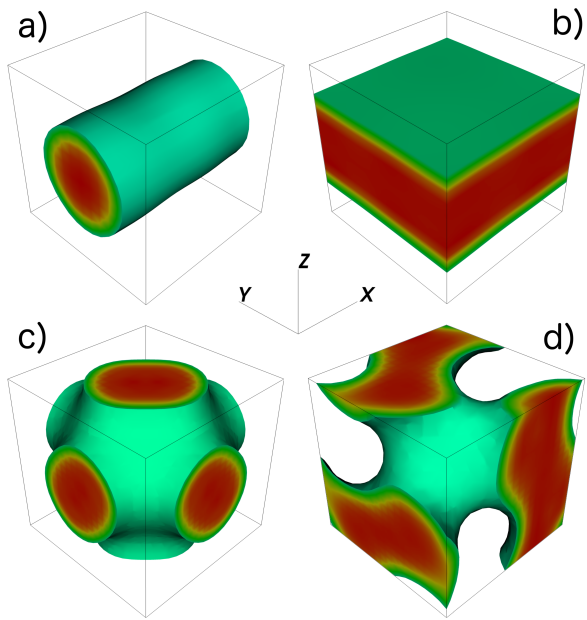}%
\caption{\label{fig:shapes}(Color online) 
Nuclear pasta shapes analyzed in
this work: (a) rod with $L=16\fm$, $A=150$; (b) slab with $L=16\fm$,
$A=294$; (c) P-surface $L=22.03\fm$, $A=762$; and (d) gyroid $L=26.01\fm$,
$A=1254$ in a cubic lattice. Dark red corresponds to $\rho=0.14 \fmd$ and green to $\rho=0.04 \fmd$
}
\end{figure}

The manuscript has been organized as follows. In Sec. \ref{sec:methods} we  describe our implementation of nuclear density functional theory and twist-averaged boundary conditions. The results of benchmarking calculations for free Fermi gas are given in Sec.~\ref{sec:results} together with results  for the rod and slab phases, and triply periodic minimal surface (TPMS) shapes: P-surface and gyroid  (see Fig.~\ref{fig:shapes}). 
We summarize our results and present the outlook
for the future in Sec.~\ref{Conclusion}.

\section{Methods}\label{sec:methods}

\subsection{Skyrme-Hartree-Fock\label{sec:HF}}

To calculate the pasta structures, we use the Skyrme-HF method, solving the Schr\"odinger equation for the many-body system in a single Slater determinant approach. For this purpose we utilize the 3D  code Sky3D \cite{Mar15a}, which solves the self-consistent HF equations with a damped gradient iteration method on an equidistant grid with no symmetry restrictions. The code, extended to the case of TABC,  uses the finite Fourier transform method for spatial derivatives. We have tested that grid spacings between 0.9\,fm and 1.05\,fm yield stable results and this is what is used in this work.

To calculate the mean field, we use the Skyrme energy density functional \cite{Bender03}:
\begin{align}
\mathcal{E}_{\rm Sk}&=\sum_{T=0,1}\left(C_T^\rho(\rho_0)\rho_T^2+
                         C_T^{\Delta\rho}\rho_T\Delta\rho_T\right.\nonumber\\
                         &+\left.C_T^{\tau}\rho_T\tau_T+
                         C_T^{\nabla\vecs{J}}\rho_T\nabla\vec{J}_T\right),\label{eq:Skyrme}
\end{align}
written in terms of the local isoscalar ($T=0$) and isovector ($T=1)$  densities and currents
$\rho_T$ (particle density), $\tau_T$ (kinetic density), and $\vec{J}_T$ (spin-orbit current).  The coupling constants of the functional correspond to the Skyrme parametrization SLy6 \cite{Chabanat}, as in our earlier work \cite{Schuetrumpf2013a,Schuetrumpf2014,Schuetrumpf2015}. The energy density functional is supplemented by the kinetic energy and  Coulomb terms.

A uniform electron background is added to ensure charge neutrality of the system. Electron screening is not included, as its influence should be very small for the  box lengths used \cite{Alcain}. We consider nuclear matter with a proton fraction of $X_P=1/3$ to facilitate comparison with earlier calculations. 

The computational cost of calculations strongly depends  on the number of nucleons and  grid size.  The main time of the calculation is consumed by the diagonalization of the HF Hamiltonian and the wave function orthonormalization. Thus, constraining the total average density is very time-consuming and small systems are preferable, at least for the purpose of benchmarking.

\subsection{Twisted Average Boundary Conditions\label{sec:TABC}}
According  to the Floquet-Bloch theorem, the single-particle (s.p.) wave functions for a particle in a periodic potential are given by:
\begin{equation}
\psi_{\alpha\vecs{q}}(\vec{r})=u_{\alpha\vecs{q}}(\vec{r})e^{\I\vecs{q}\vecs{r}}\ ,
\end{equation}
where $\alpha$ is a discrete label of the wave function, $\vec{q}$ is the wave vector that determines the boundary condition, and $ u_{\alpha\vecs{q}}(\vec{r}) $ is a periodic function of $\vec{r}$. The wave vector can  be replaced by three twist angles $\theta_i=\vec{q}\vec{T}_i ~(i=x,y,z)$, where $\vec{T}_i=L_i\vec{e}_i$ are the lattice vectors with the unit vectors $\vec{e}_i$. The Bloch waves  corresponding to PBC represent a particular case of $\vec{q}=\vec{\theta}=0$.

The general Bloch boundary conditions can be written as:
\begin{equation}\label{bbc}
\psi_{\alpha\vecs{\theta}}(\vec{r}+\vec{T}_i)=e^{\I\theta_i}\psi_{\alpha\vecs{\theta}}(\vec{r}).
\end{equation}
The s.p.wave functions $\psi_{\alpha\vecs{\theta}}(\vec{r})$ defining the HF densities and fields  are eigenstates of of the
HF Hamiltonian $\hat{h}_{\vecs{\theta}}$ corresponding to the boundary condition (\ref{bbc}):
\begin{equation}
\hat{h}_{\vecs{\theta}}\psi_{\alpha\vecs{\theta}}(\vec{r})=\epsilon_{\alpha\vecs{\theta}}\psi_{\alpha\vecs{\theta}}(\vec{r}).
\end{equation}

In the TABC method, the expectation value of an observable $\hat{O}$ is obtained by
 averaging over the twist angles:
\begin{equation}
\langle \hat{O}\rangle= \int  \frac{d^3\vec{\theta}}{\pi^3}\,\langle \Psi_{\vecs{\theta}}| \hat{O}| \Psi_{\vecs{\theta}}\rangle,
\label{eq:average}
\end{equation}
where $\Psi_{\vecs{\theta}}$ is the HF product wave function. In Eq.~(\ref{eq:average}),  the angles
$ \theta_i $ change  between zero (PBC) and $\pi$ (anti-PBC), as the time-reversal symmetry is assumed \cite{Lin}. In practical calculations discussed in this study, the integrals over the Bloch phases are computed by means of Gauss-Legendre (GL) quadrature  with $N_{\rm GL}$=4 points, unless stated otherwise.

\section{Results}\label{sec:results}

\subsection{Free particle gas\label{sec:free}}

\begin{figure}[tb]
\includegraphics[width=0.9\linewidth]{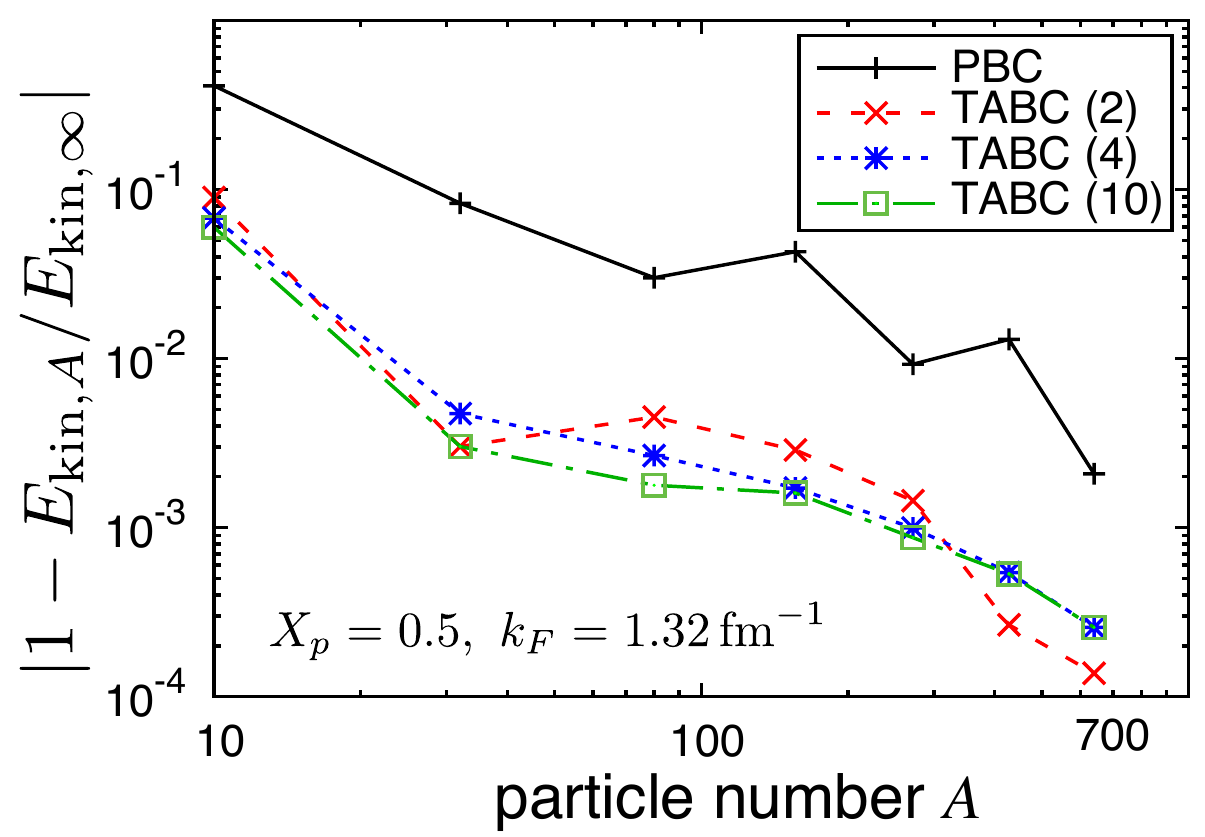}%
\caption{\label{fig:free}(Color online) Relative finite-size correction for the kinetic energy of free symmetric nucleonic matter versus the total nucleon number $A$ at $k_F\approx1.32\fmk$ computed with PBS and TABS. The TABS averaging was carried out with  2, 4, and 10 GL quadrature points in each direction.}
\end{figure}

Following Ref.~\cite{Hagen14}, to test  our TABC implementation, we consider  a gas of non-interacting neutrons and protons with $X_P=0.5$. The exact solution for the average kinetic energy can be easily derived  from the Fermi gas model. Here, we take a system with a total density of $\rho=0.15625\fmd$ ($k_F\approx1.32\fmk$) and an average kinetic energy of $E_{\rm kin, \infty}\approx21.7786\MeV$.
For plane waves in a cubic box $L^3$, the resulting wave numbers are quantized:
\begin{equation}
k_{i,n}=\frac{2\pi n+\theta_i}{L_i}\quad n=0,\pm1,\pm 2, ... ,\pm n_{\rm max}.
\end{equation}
By averaging over $\vec{\theta}$, we recover a continuous spectrum of $\vec{k}$; hence,  the infinite-volume limit should be reached much faster. 

The results are shown in Fig.~\ref{fig:free}. The finite-volume corrections for different system sizes are plotted for PBC and TABS with 2-, 4-, and 10-point GL averaging. The results show that the finite-size fluctuations are reduced drastically, even with a very small number of GL integration points. With a higher number of GL points, a slightly smoother convergence can be reached.
In general, with the TABC method,  one gains an order of magnitude  in precision in this case ~\cite{Hagen14}. 

The number of calculations quickly grows as  $N_{GL}^d$, where $d$ is the dimension of the phase. For majority of  pasta phases, it can be assumed that permutations of $\theta_i$ give the same result due to symmetry considerations. For $d=3$ the number of calculations decreases to $(2N_{\rm GL}^3+6N_{\rm GL}^2+4N_{\rm GL})/12$. For the following pasta calculations we set $N_{\rm GL}=4$. This means, that we perform 20 calculations for each system with $d=3$.

\subsection{Pasta phases\label{sec:rodslab}}

In the following, we apply TABC to  actual pasta phases. To ensure that the calculations converge at desired shapes, we add a guiding potential $\phi_p$ during the first 200 HF iterations~\cite{Schuetrumpf2015}:
\begin{subequations}
  \begin{align}
    \phi_{\rm R} &=-\phi_0\left(\cos\,Y+\cos\,Z\right), \\
    \phi_{\rm S} &=-\phi_0\left(\cos Z\right),\\
    \phi_{\rm P} &=\phi_0\left(\cos X+\cos Y+\cos Z\right), \label{eq:nodalP} \\
    \phi_{\rm G} &=\phi_0\left(\cos X \sin Y + \cos Y \sin Z + \cos Z \sin X\right), \label{eq:nodalG} 
  \end{align}
\end{subequations}
where $p \in \{{\rm R,S,P,G}\}$ for rod, slab, P-surface, and gyroid, respectively,
and $X_i=2\pi x_i/L_i$. The potentials for P and G shapes are their first-order nodal approximations \cite{NesperSchnering1991}. The parameter $\phi_0>0$ has to be chosen such that the guiding potential is similar to the resulting self consistent potential to guarantee a stable and fast convergence.

\subsubsection{Rod}
\begin{figure}[htb]
\includegraphics[width=0.95\linewidth]{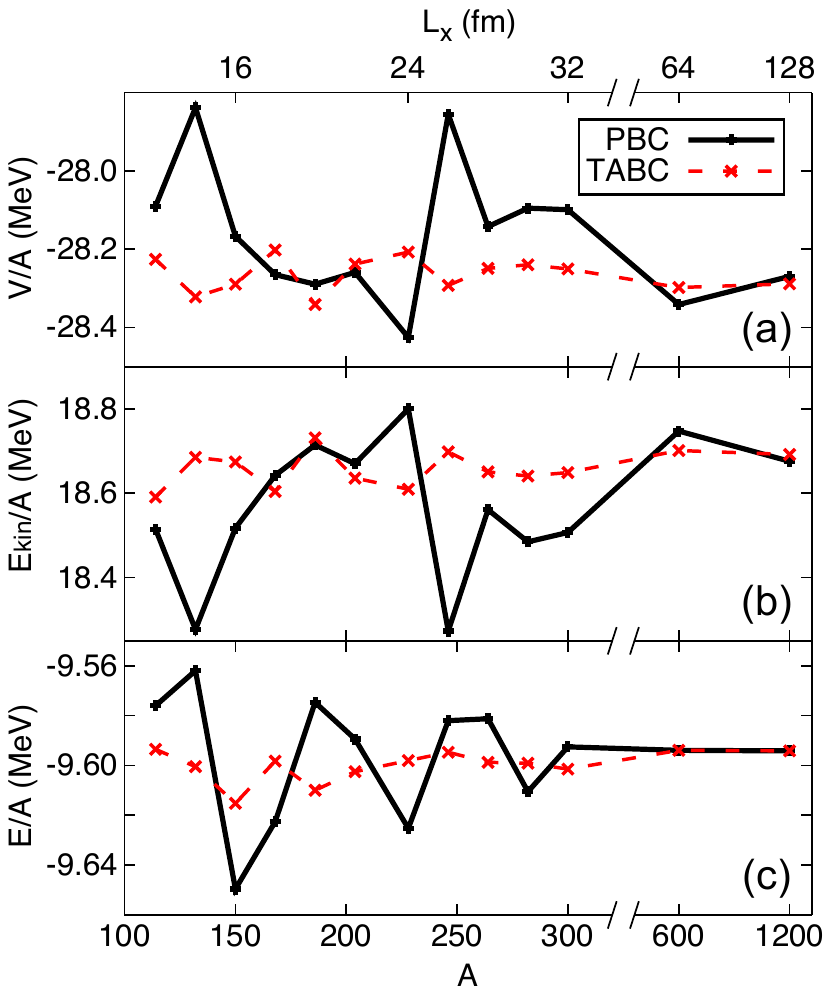}%
\caption{\label{fig:rod}(Color online) The total energy $E/A$ (bottom),  kinetic energy $E_{\rm kin}/A$ (middle), and  potential energy $V/A$ (top) per particle for the rod phase of Fig.~\ref{fig:shapes}(a) 
at $\rho=0.0358\fmd$ computed with  PBC (solid line) and TABC (dashed line) as a function of particle number $A$ (or the box length $L_x$).}
\end{figure}

The first example concerns the rod phase. The corresponding shape shown in Fig.~\ref{fig:shapes}(a) is axially symmetric around the $x$-axis. We calculate one rod in a rectangular box with $L_y=L_z=16\,\fm$. Note that the lowest-energy rod configuration corresponds to a honeycomb arrangement~\cite{Oka13a}, but the single-rod configuration considered here serves well as an illustration of TABC. 
In this example, we vary the particle number $A$ and the box length $L_x$ simultaneously to maintain the density  and  the distance from the rods in the neighboring cells. The results of our calculations are shown in Fig.~\ref{fig:rod} in the range of $100 \leq A \leq 1200$.

In the limit of large particle number (or $L_x$), the energy per nucleon should be constant, and this limit is reached for particle numbers below  $A=1200$. 
In the PBC variant, the magnitude of finite-size corrections manifesting themselves as energy fluctuations in   Fig.~\ref{fig:rod} is large; it reaches $\approx0.04\MeV/A$. It is gratifying to see that the fluctuations are significantly reduced in TABC. For $A>200$ the magnitude of finite-size effects falls below $0.01\MeV/A$. For the   potential and kinetic energies
shown in  Figs.~\ref{fig:rod}(a) and \ref{fig:rod}(b), respectively, the spurious fluctuations are larger; here the TABC method reduces the finite-size effects below $0.1\MeV/A$, while they are as large as $0.4\MeV/A$ in the PBC variant.

\subsubsection{Slab}

A similar test can be done for the slab phase of Fig.~\ref{fig:shapes}(b). This shape is translationally invariant along $x$ and $y$  directions. In the first set of calculations we keep the box length in the $z$-direction, $L_z=16\fm$, and vary the box length in the perpendicular directions simultaneously with the particle number to maintain the constant density  $\rho=0.0715\fmd$. At this density, the slab is confined to approximately half the volume of the box \cite{Schuetrumpf2015}.

\begin{figure}[htb]
\includegraphics[width=0.95\linewidth]{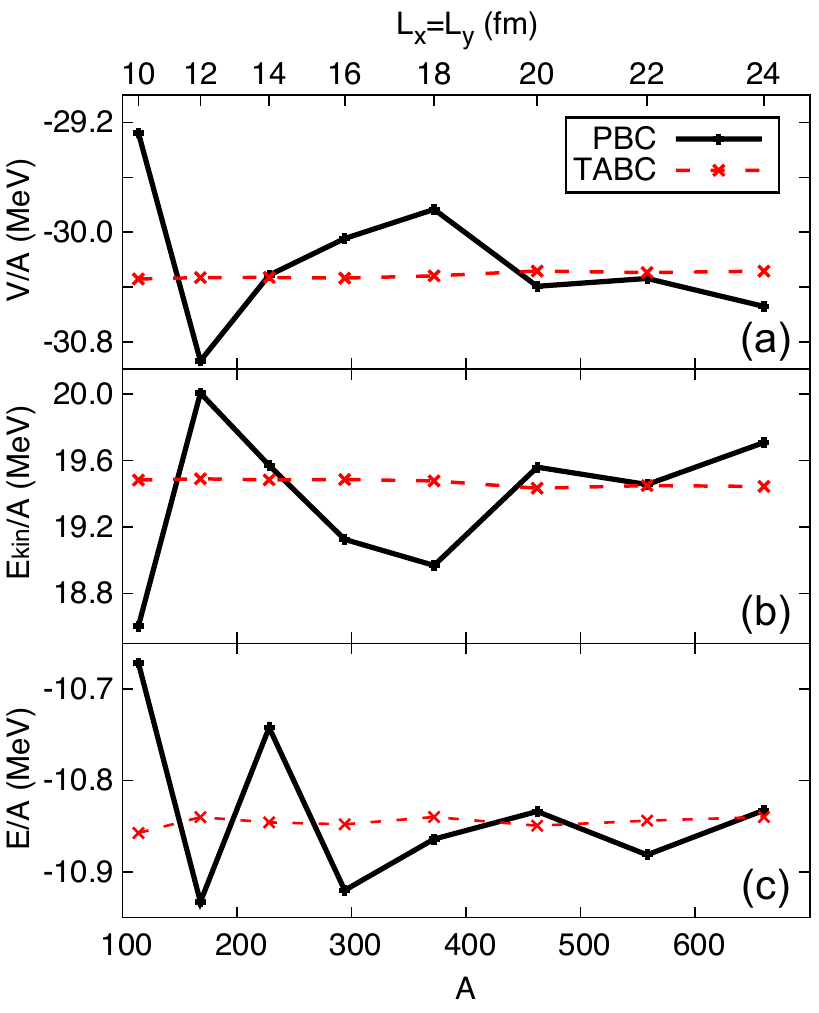}%
\caption{\label{fig:slab}(Color online) Similar as in Fig.~\ref{fig:rod} but for the slab phase of Fig.~\ref{fig:shapes}(b) at $\rho=0.0715\fmd$ as a function of $A$ (or $L_x=L_y$).}
\end{figure}

The results of our test calculations are shown in Fig.~\ref{fig:slab}. Here the improvement provided by TABC is even more impressive than for the rod shape, because the 2-dimensional averaging is more effective. The plateau in $E/A$ is reached well below $A=660$ in TABC.  Improvements for $T/A$ and $V/A$ are also significant: the range of fluctuations
is reduced from $\approx1.1\MeV/A$ in PBC to
 $\approx0.056\MeV/A$ in TABC.

\begin{figure}[htb]
\includegraphics[width=0.95\linewidth]{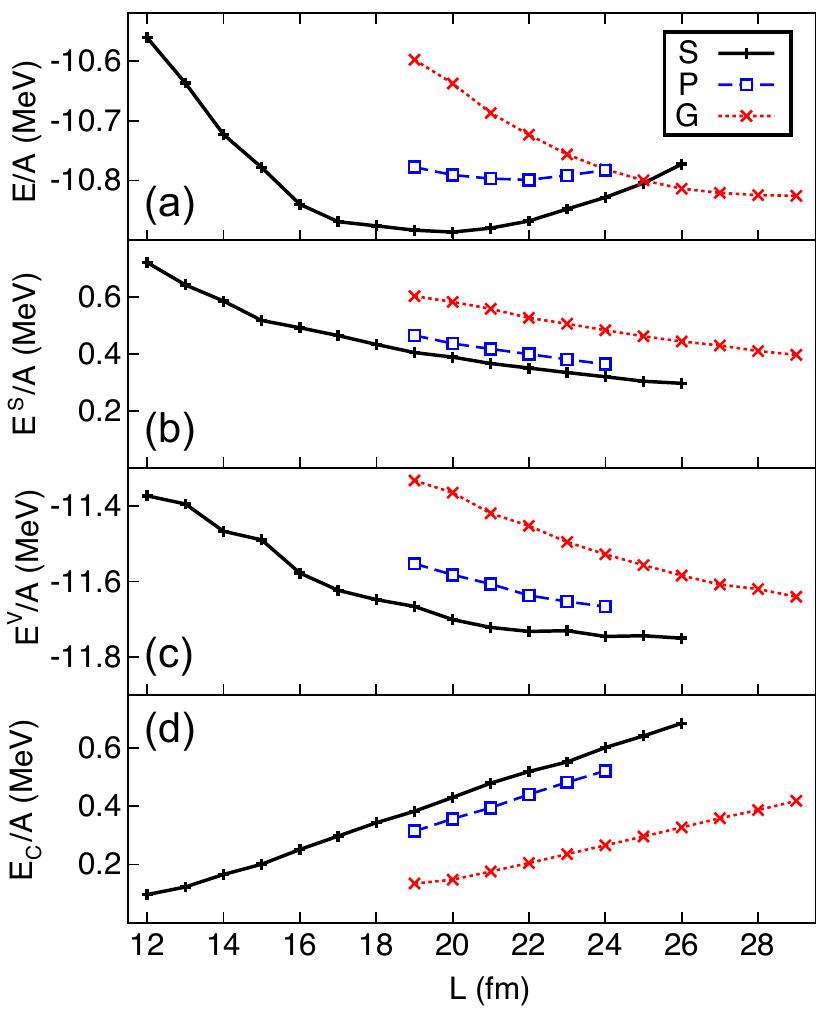}%
\caption{\label{fig:spg-opt}(Color online) TABC energies per nucleon versus the periodic length $L$ of the slab shape (solid line), P-surface (dashed line) and Gyroid (dotted line) at $\rho=0.0715\fmd$. 
Shown are: the total energy (a); surface energy (b); volume energy (c); and the Coulomb energy (d).}
\end{figure}

Another TABC benchmark for the slab phase can be obtained by varying the box length $L_z$ while keeping perpendicular lengths constant, $L_x=L_y=16$\,fm. Here, the particle number is adjusted to maintain the thickness of the slab at $\approx L_z/2$. This has already been done in the PBC calculations of Ref.~\cite{Schuetrumpf2015} but the magnitude of finite-size fluctuations turned out to be  so large that the trend of the total energy with $A$ was impossible to assess. The results of TABC are shown in Fig.~\ref{fig:spg-opt} (solid line).

The results for the total energy are shown in Fig.~\ref{fig:spg-opt}(a). A clear minimum at $L_z\approx20\fm$ and  $E/A=-10.89\MeV$ is found. The reason for this minimum can be understood by inspecting  different contributions to the total energy. To this end, the total energy has been decomposed  into three parts: 
the volume energy  $E^V$, the surface energy  $E^S$, and  
the Coulomb energy $E_C$. The volume and surface terms are defined as:
\begin{eqnarray}\label{eq:SkyrmeBS}
E^V&=&E_{\rm kin} + \sum_{T=0,1} E^V_{{\rm Sk},T}, \label{EV}\\
E^S&=&\sum_{T=0,1} E^S_{{\rm Sk},T}, \label{ES}
\end{eqnarray}
where
\begin{subequations}
 \begin{eqnarray}
E^V_{{\rm Sk},T}&=\int d^3r \left(C_T^\rho(\rho_0)\rho_T^2+
                         C_T^{\tau}\rho_T\tau_T\right), \label{EVT}\\
E^S_{{\rm Sk},T}&=\int  d^3r \left(C_T^{\Delta\rho}\rho_T\Delta\rho_T+
                         C_T^{\nabla\vecs{J}}\rho_T\nabla\vec{J}_T\right) \label{EST}.
 \end{eqnarray}
\end{subequations}

As seen in Fig.~\ref{fig:spg-opt}(d), for the small slabs the Coulomb energy is very low, because the positively charged slabs and the negatively charged voids where the electron gas dominates are located very close to each other, and this results in a cancellation between electrostatic repulsion and attraction.
For large slabs, however,   the Coulomb repulsion dominates. It appears that this results in an almost linear behavior of $E_C$ as a function of $L$.

The volume energy shows a decreasing trend with $L$; it flattens out for large box lengths. At the limit of $L\rightarrow\infty$ the volume energy is supposed to  reach the value of  $E_\infty^V\approx-12.4\MeV$ ($E/A$ for nuclear matter with $X_P=1/3$ at the equilibrium density of $0.147\fmd$). As illustrated in Fig.~\ref{fig:spg-opt}(c), $E^V$ is not close to this value at $L=26\fm$, because the density within the slab is not yet constant and surface effects are still  important.

The surface energy per nucleon in Fig.~\ref{fig:spg-opt}(b) decreases as $L_z^{-1}=A^{-1}$ as the surface area $2L_xL_y$ is kept constant and the particle number is proportional to $L_z$ (as the particle density is fixed). It is seen that  the surface energy provides an appreciable contribution to the total energy even for the largest box lengths probed in our calculations.
For small slabs,  volume and surface  terms dominate the behavior of $E/A$, as both contributions decrease rapidly with $L$. For large box lengths, the pattern of $E/A$ is dominated by the Coulomb effect. 

\subsubsection{TPMS\label{sec:TPMS}}

At the same mid-density range where the slab phases appear,  other pasta phases are predicted as well. Those are triply periodic minimal surface phases, or TPMS. For TPMS shapes, TABC works even better than for the slabs, as three-dimensional averaging can be performed. Here we vary $L_x=L_y=L_z$ simultaneously to maintain a cubic box. The specific shapes analyzed in this work are the P-surface shape and the gyroid shown in Figs.~\ref{fig:shapes}(c) and \ref{fig:shapes}(d), respectively. Of particular relevance is the gyroid phase  \cite{Nakazato2009,Nakazato2011}. Minimal surfaces are of considerable interest because of their vanishing mean curvature and negative Gaussian curvature. They have been found in multiple soft-matter systems \cite{MichielsenStavenga:2008, SaranathanOsujiMochrieNohNarayananSandyDufresnePrum:2010, SchroederTurkWickhamAverdunkBrinkFitzGeraldPoladianLargeHyde:2011, Hajduk1994, Larsson:1989}. The nodal approximation of the P-surface and the gyroid are shown in Eqs. (\ref{eq:nodalP}) and (\ref{eq:nodalG}) setting the right-hand side to zero. These structures were predicted to appear in nuclear matter in time-dependent HF simulations  \cite{Schuetrumpf2015}. 

Furthermore, the P-surface, the gyroid surface, and also the D-Surface (diamond; not discussed here), are closely related via the Bonnet transformation \cite{HydeLanguageOfShape:1997}. Keeping the surface isometric, one can derive the relationship between the unit cell lattice parameters: $L_P/L_G=0.812$. The ratio of the surface area for the same lattice parameters is $A_P/A_G=0.758$; this can be compared  to the slab in the same cubic box: $A_P/A_S=1.172$  \cite{Schroeder2003}. For both  P- and G-structures, the volume occupied by nuclear matter is half of the total volume. Moreover, P, D, and G are optimal structures with minimal variations of the Gaussian curvature, and G is the optimal structure with minimal variations of the structure width \cite{FogdenHyde:1999,Schroeder2003,SchroederFogdenHyde:2006}.

The TABC results for the P-surface and the gyroid are shown  in Fig.~\ref{fig:spg-opt}. At low values of $L$,  gyroids and P-surfaces are not stable at certain ranges of twist angles. For that reason, we present results only for $L\geq 19$\,fm. The largest calculations for the gyroids in the $29\fm$ box correspond to  1734 particles. This is close to the  limit of our HF solver Sky3D. 

The minimum of the total energy per nucleon is reached at a box length of $21.65\fm$ for the P-surface and at around $28\fm$ for the gyroid, see
Fig.~\ref{fig:spg-opt}(a). The surface energies in Fig.~\ref{fig:spg-opt}(b) reflect the ratios of the surface areas for gyroid, P-surface, and slab discussed above. Similar to the slab case discussed earlier, the surface energy per particle for P- and G-shapes decreases as $L^{-1}$, because  the particle number is proportional to $L^3$ while the surface area grows as $L^2$. 
 
Compared to the slab case,  the Coulomb energy is lower for the TPMS at a given box size, because of less compact distributions. While for the P-surface this is  a minor effect, the Coulomb energy for the gyroid is reduced by a factor of 
$\sim$2. This is reflected in the behavior of the total energy in Fig.~\ref{fig:spg-opt}(a): the minima of TPMS are shifted to higher values of  $L$. However, the electrostatic repulsion is better compensated by  the nuclear energy for TPMS; hence, the energy minima for P- and G-shapes are not as deep as in the slab case. It should be noted that the TPMS minima, especially for the gyroid, are close to the energy minimum of the slab phase.

\section{Conclusions\label{Conclusion}}
We implemented the TABC approach into the 3D HF framework used to simulate pasta phases in the neutron star crust. We demonstrated that by averaging over Bloch boundary conditions one is able to significantly reduce the magnitude of spurious finite-volume effects for Bloch-Wigner cells containing hundreds of nucleons.

Practical calculations were carried out for asymmetric matter with $X_p=0.3$. We first benchmarked TABC for the nucleonic gas. The results turned out to be weakly sensitive to the number of integration points. We found that taking  $N_{\rm GL}=4$ Gauss-Legendre points yields stable results. 
The TABC method was then  tested for the rod and slab phases, simultaneously varying the box lengths and the number of particles to keep the average density constant. For the rod phase, the finite-volume error was reduced by a factor of more than three, down  to $\Delta E=0.02\MeV/A$. For the two-dimensional slab shapes the improvements are even more significant.

By eliminating spurious finite-size fluctuations through TABC, we were able to inspect individual energy  contributions from various terms of the energy density functional.  This was done by varying the physical length of the box. We showed that the energy variation primarily comes from the Coulomb and the surface terms. We have also demonstrated that, due to its unique geometry, the gyroid geometry minimizes the Coulomb energy drastically for a given box length as compared to the slab and the P-surface phases.

Future applications will include the TABC extension of the adaptive multi-resolution 3D Hartree-Fock solver \cite{Pei2014} and Hartree-Fock-Bogoliubov 
TABC applications to superfluid pasta phases and complex nucleonic topologies as in fission. The lesson learned form the exercise presented in this study is that high-fidelity results for pasta phases can be obtained by considering finite-volume boxes containing up to several thousand particles. Moreover, TABC calculations  are very well suited to parallel  computing as HF computations
at different twist angles can be carried out independently. 

\begin{acknowledgements}
This material is based upon work supported by the U.S. Department of Energy, Office of Science under Award Numbers
DOE-DE-NA0002574 (the Stewardship Science Academic Alliances
program) and   DE-SC0008511 (NUCLEI SciDAC-3 collaboration)
This work used computational resources of  the Institute for Cyber-Enabled Research at  Michigan State University. We are grateful to P.-G. Reinhard for useful discussions.
\end{acknowledgements}

\bibliography{TABC}

\begin{thebibliography}{53}%
\makeatletter
\providecommand \@ifxundefined [1]{%
 \@ifx{#1\undefined}
}%
\providecommand \@ifnum [1]{%
 \ifnum #1\expandafter \@firstoftwo
 \else \expandafter \@secondoftwo
 \fi
}%
\providecommand \@ifx [1]{%
 \ifx #1\expandafter \@firstoftwo
 \else \expandafter \@secondoftwo
 \fi
}%
\providecommand \natexlab [1]{#1}%
\providecommand \enquote  [1]{``#1''}%
\providecommand \bibnamefont  [1]{#1}%
\providecommand \bibfnamefont [1]{#1}%
\providecommand \citenamefont [1]{#1}%
\providecommand \href@noop [0]{\@secondoftwo}%
\providecommand \href [0]{\begingroup \@sanitize@url \@href}%
\providecommand \@href[1]{\@@startlink{#1}\@@href}%
\providecommand \@@href[1]{\endgroup#1\@@endlink}%
\providecommand \@sanitize@url [0]{\catcode `\\12\catcode `\$12\catcode
  `\&12\catcode `\#12\catcode `\^12\catcode `\_12\catcode `\%12\relax}%
\providecommand \@@startlink[1]{}%
\providecommand \@@endlink[0]{}%
\providecommand \url  [0]{\begingroup\@sanitize@url \@url }%
\providecommand \@url [1]{\endgroup\@href {#1}{\urlprefix }}%
\providecommand \urlprefix  [0]{URL }%
\providecommand \Eprint [0]{\href }%
\providecommand \doibase [0]{http://dx.doi.org/}%
\providecommand \selectlanguage [0]{\@gobble}%
\providecommand \bibinfo  [0]{\@secondoftwo}%
\providecommand \bibfield  [0]{\@secondoftwo}%
\providecommand \translation [1]{[#1]}%
\providecommand \BibitemOpen [0]{}%
\providecommand \bibitemStop [0]{}%
\providecommand \bibitemNoStop [0]{.\EOS\space}%
\providecommand \EOS [0]{\spacefactor3000\relax}%
\providecommand \BibitemShut  [1]{\csname bibitem#1\endcsname}%
\let\auto@bib@innerbib\@empty
\bibitem [{\citenamefont {Ravenhall}\ \emph {et~al.}(1983)\citenamefont
  {Ravenhall}, \citenamefont {Pethick},\ and\ \citenamefont
  {Wilson}}]{Ravenhall}%
  \BibitemOpen
  \bibfield  {author} {\bibinfo {author} {\bibfnamefont {D.~G.}\ \bibnamefont
  {Ravenhall}}, \bibinfo {author} {\bibfnamefont {C.~J.}\ \bibnamefont
  {Pethick}}, \ and\ \bibinfo {author} {\bibfnamefont {J.~R.}\ \bibnamefont
  {Wilson}},\ }\href {\doibase 10.1103/PhysRevLett.50.2066} {\bibfield
  {journal} {\bibinfo  {journal} {Phys. Rev. Lett.}\ }\textbf {\bibinfo
  {volume} {50}},\ \bibinfo {pages} {2066} (\bibinfo {year}
  {1983})}\BibitemShut {NoStop}%
\bibitem [{\citenamefont {Hashimoto}\ \emph {et~al.}(1984)\citenamefont
  {Hashimoto}, \citenamefont {Seki},\ and\ \citenamefont {Yamada}}]{Hashimoto}%
  \BibitemOpen
  \bibfield  {author} {\bibinfo {author} {\bibfnamefont {M.}~\bibnamefont
  {Hashimoto}}, \bibinfo {author} {\bibfnamefont {H.}~\bibnamefont {Seki}}, \
  and\ \bibinfo {author} {\bibfnamefont {M.}~\bibnamefont {Yamada}},\ }\href
  {\doibase 10.1143/PTP.71.320} {\bibfield  {journal} {\bibinfo  {journal}
  {Prog. Theor. Phys.}\ }\textbf {\bibinfo {volume} {71}},\ \bibinfo {pages}
  {320} (\bibinfo {year} {1984})}\BibitemShut {NoStop}%
\bibitem [{\citenamefont {Horowitz}\ \emph
  {et~al.}(2004{\natexlab{a}})\citenamefont {Horowitz}, \citenamefont
  {P\'erez-Garc\'ia},\ and\ \citenamefont {Piekarewicz}}]{Horowitz2004}%
  \BibitemOpen
  \bibfield  {author} {\bibinfo {author} {\bibfnamefont {C.~J.}\ \bibnamefont
  {Horowitz}}, \bibinfo {author} {\bibfnamefont {M.~A.}\ \bibnamefont
  {P\'erez-Garc\'ia}}, \ and\ \bibinfo {author} {\bibfnamefont
  {J.}~\bibnamefont {Piekarewicz}},\ }\href {\doibase
  10.1103/PhysRevC.69.045804} {\bibfield  {journal} {\bibinfo  {journal} {Phys.
  Rev. C}\ }\textbf {\bibinfo {volume} {69}},\ \bibinfo {pages} {045804}
  (\bibinfo {year} {2004}{\natexlab{a}})}\BibitemShut {NoStop}%
\bibitem [{\citenamefont {Horowitz}\ \emph
  {et~al.}(2004{\natexlab{b}})\citenamefont {Horowitz} \emph
  {et~al.}}]{Horowitz20042}%
  \BibitemOpen
  \bibfield  {author} {\bibinfo {author} {\bibfnamefont {C.~J.}\ \bibnamefont
  {Horowitz}} \emph {et~al.},\ }\href {\doibase 10.1103/PhysRevC.70.065806}
  {\bibfield  {journal} {\bibinfo  {journal} {Phys. Rev. C}\ }\textbf {\bibinfo
  {volume} {70}},\ \bibinfo {pages} {065806} (\bibinfo {year}
  {2004}{\natexlab{b}})}\BibitemShut {NoStop}%
\bibitem [{\citenamefont {Sonoda}\ \emph {et~al.}(2007)\citenamefont {Sonoda}
  \emph {et~al.}}]{Sonoda2007}%
  \BibitemOpen
  \bibfield  {author} {\bibinfo {author} {\bibfnamefont {H.}~\bibnamefont
  {Sonoda}} \emph {et~al.},\ }\href {\doibase 10.1103/PhysRevC.75.042801}
  {\bibfield  {journal} {\bibinfo  {journal} {Phys. Rev. C}\ }\textbf {\bibinfo
  {volume} {75}},\ \bibinfo {pages} {042801} (\bibinfo {year}
  {2007})}\BibitemShut {NoStop}%
\bibitem [{\citenamefont {Grygorov}\ \emph {et~al.}(2010)\citenamefont
  {Grygorov}, \citenamefont {G{\"o}gelein},\ and\ \citenamefont
  {M{\"u}ther}}]{Gry10}%
  \BibitemOpen
  \bibfield  {author} {\bibinfo {author} {\bibfnamefont {P.}~\bibnamefont
  {Grygorov}}, \bibinfo {author} {\bibfnamefont {P.}~\bibnamefont
  {G{\"o}gelein}}, \ and\ \bibinfo {author} {\bibfnamefont {H.}~\bibnamefont
  {M{\"u}ther}},\ }\href {http://stacks.iop.org/0954-3899/37/i=7/a=075203}
  {\bibfield  {journal} {\bibinfo  {journal} {J. Phys. G}\ }\textbf {\bibinfo
  {volume} {37}},\ \bibinfo {pages} {075203} (\bibinfo {year}
  {2010})}\BibitemShut {NoStop}%
\bibitem [{\citenamefont {Gusakov}\ \emph {et~al.}(2004)\citenamefont
  {Gusakov}, \citenamefont {Yankovlev}, \citenamefont {P.Haensel},\ and\
  \citenamefont {Gnedin}}]{Gusakov}%
  \BibitemOpen
  \bibfield  {author} {\bibinfo {author} {\bibfnamefont {M.~E.}\ \bibnamefont
  {Gusakov}}, \bibinfo {author} {\bibfnamefont {D.~G.}\ \bibnamefont
  {Yankovlev}}, \bibinfo {author} {\bibnamefont {P.Haensel}}, \ and\ \bibinfo
  {author} {\bibfnamefont {O.~Y.}\ \bibnamefont {Gnedin}},\ }\href@noop {}
  {\bibfield  {journal} {\bibinfo  {journal} {Astron. Astrophys.}\ }\textbf
  {\bibinfo {volume} {412}},\ \bibinfo {pages} {1143} (\bibinfo {year}
  {2004})}\BibitemShut {NoStop}%
\bibitem [{\citenamefont {L.Bildsten}\ and\ \citenamefont
  {Ushomirsky}(2000)}]{Bildsten}%
  \BibitemOpen
  \bibfield  {author} {\bibinfo {author} {\bibnamefont {L.Bildsten}}\ and\
  \bibinfo {author} {\bibfnamefont {G.}~\bibnamefont {Ushomirsky}},\
  }\href@noop {} {\bibfield  {journal} {\bibinfo  {journal} {Astrophys. J.
  Lett.}\ }\textbf {\bibinfo {volume} {529}},\ \bibinfo {pages} {L33} (\bibinfo
  {year} {2000})}\BibitemShut {NoStop}%
\bibitem [{\citenamefont {Lindblom}\ \emph {et~al.}(2000)\citenamefont
  {Lindblom}, \citenamefont {Owen},\ and\ \citenamefont
  {Ushomirsky}}]{Lindblom00}%
  \BibitemOpen
  \bibfield  {author} {\bibinfo {author} {\bibfnamefont {L.}~\bibnamefont
  {Lindblom}}, \bibinfo {author} {\bibfnamefont {B.~J.}\ \bibnamefont {Owen}},
  \ and\ \bibinfo {author} {\bibfnamefont {G.}~\bibnamefont {Ushomirsky}},\
  }\href@noop {} {\bibfield  {journal} {\bibinfo  {journal} {Phys. Rev. D}\
  }\textbf {\bibinfo {volume} {62}},\ \bibinfo {pages} {084030} (\bibinfo
  {year} {2000})}\BibitemShut {NoStop}%
\bibitem [{\citenamefont {Dorso}\ \emph {et~al.}(2012)\citenamefont {Dorso},
  \citenamefont {Gim\'enez~Molinelli},\ and\ \citenamefont
  {L\'opez}}]{Dorso2012}%
  \BibitemOpen
  \bibfield  {author} {\bibinfo {author} {\bibfnamefont {C.~O.}\ \bibnamefont
  {Dorso}}, \bibinfo {author} {\bibfnamefont {P.~A.}\ \bibnamefont
  {Gim\'enez~Molinelli}}, \ and\ \bibinfo {author} {\bibfnamefont {J.~A.}\
  \bibnamefont {L\'opez}},\ }\href {\doibase 10.1103/PhysRevC.86.055805}
  {\bibfield  {journal} {\bibinfo  {journal} {Phys. Rev. C}\ }\textbf {\bibinfo
  {volume} {86}},\ \bibinfo {pages} {055805} (\bibinfo {year}
  {2012})}\BibitemShut {NoStop}%
\bibitem [{\citenamefont {Williams}\ and\ \citenamefont
  {Koonin}(1985)}]{Williams1985}%
  \BibitemOpen
  \bibfield  {author} {\bibinfo {author} {\bibfnamefont {R.}~\bibnamefont
  {Williams}}\ and\ \bibinfo {author} {\bibfnamefont {S.}~\bibnamefont
  {Koonin}},\ }\href {\doibase 10.1016/0375-9474(85)90191-5} {\bibfield
  {journal} {\bibinfo  {journal} {Nucl. Phys.}\ }\textbf {\bibinfo {volume}
  {435}},\ \bibinfo {pages} {844 } (\bibinfo {year} {1985})}\BibitemShut
  {NoStop}%
\bibitem [{\citenamefont {Okamoto}\ \emph {et~al.}(2013)\citenamefont
  {Okamoto}, \citenamefont {Maruyama}, \citenamefont {Yabana},\ and\
  \citenamefont {Tatsumi}}]{Oka13a}%
  \BibitemOpen
  \bibfield  {author} {\bibinfo {author} {\bibfnamefont {M.}~\bibnamefont
  {Okamoto}}, \bibinfo {author} {\bibfnamefont {T.}~\bibnamefont {Maruyama}},
  \bibinfo {author} {\bibfnamefont {K.}~\bibnamefont {Yabana}}, \ and\ \bibinfo
  {author} {\bibfnamefont {T.}~\bibnamefont {Tatsumi}},\ }\href@noop {}
  {\bibfield  {journal} {\bibinfo  {journal} {Phys. Rev. C}\ }\textbf {\bibinfo
  {volume} {88}},\ \bibinfo {pages} {025801} (\bibinfo {year}
  {2013})}\BibitemShut {NoStop}%
\bibitem [{\citenamefont {Pais}\ \emph {et~al.}(2015)\citenamefont {Pais},
  \citenamefont {Chiacchiera},\ and\ \citenamefont {Provid{\^e}ncia}}]{Pais15}%
  \BibitemOpen
  \bibfield  {author} {\bibinfo {author} {\bibfnamefont {H.}~\bibnamefont
  {Pais}}, \bibinfo {author} {\bibfnamefont {S.}~\bibnamefont {Chiacchiera}}, \
  and\ \bibinfo {author} {\bibfnamefont {C.}~\bibnamefont {Provid{\^e}ncia}},\
  }\href {\doibase 10.1103/PhysRevC.91.055801} {\bibfield  {journal} {\bibinfo
  {journal} {Phys. Rev. C}\ }\textbf {\bibinfo {volume} {91}},\ \bibinfo
  {pages} {055801} (\bibinfo {year} {2015})}\BibitemShut {NoStop}%
\bibitem [{\citenamefont {Schneider}\ \emph {et~al.}(2013)\citenamefont
  {Schneider}, \citenamefont {Horowitz}, \citenamefont {Hughto},\ and\
  \citenamefont {Berry}}]{Schneider13}%
  \BibitemOpen
  \bibfield  {author} {\bibinfo {author} {\bibfnamefont {A.~S.}\ \bibnamefont
  {Schneider}}, \bibinfo {author} {\bibfnamefont {C.~J.}\ \bibnamefont
  {Horowitz}}, \bibinfo {author} {\bibfnamefont {J.}~\bibnamefont {Hughto}}, \
  and\ \bibinfo {author} {\bibfnamefont {D.~K.}\ \bibnamefont {Berry}},\ }\href
  {\doibase 10.1103/PhysRevC.88.065807} {\bibfield  {journal} {\bibinfo
  {journal} {Phys. Rev. C}\ }\textbf {\bibinfo {volume} {88}},\ \bibinfo
  {pages} {065807} (\bibinfo {year} {2013})}\BibitemShut {NoStop}%
\bibitem [{\citenamefont {Schneider}\ \emph {et~al.}(2014)\citenamefont
  {Schneider}, \citenamefont {Berry}, \citenamefont {Briggs}, \citenamefont
  {Caplan},\ and\ \citenamefont {Horowitz}}]{Schneider14}%
  \BibitemOpen
  \bibfield  {author} {\bibinfo {author} {\bibfnamefont {A.~S.}\ \bibnamefont
  {Schneider}}, \bibinfo {author} {\bibfnamefont {D.~K.}\ \bibnamefont
  {Berry}}, \bibinfo {author} {\bibfnamefont {C.~M.}\ \bibnamefont {Briggs}},
  \bibinfo {author} {\bibfnamefont {M.~E.}\ \bibnamefont {Caplan}}, \ and\
  \bibinfo {author} {\bibfnamefont {C.~J.}\ \bibnamefont {Horowitz}},\
  }\href@noop {} {\bibfield  {journal} {\bibinfo  {journal} {Phys. Rev. C}\
  }\textbf {\bibinfo {volume} {90}},\ \bibinfo {pages} {055805} (\bibinfo
  {year} {2014})}\BibitemShut {NoStop}%
\bibitem [{\citenamefont {Maruyama}\ \emph {et~al.}(1998)\citenamefont
  {Maruyama} \emph {et~al.}}]{Maruyama}%
  \BibitemOpen
  \bibfield  {author} {\bibinfo {author} {\bibfnamefont {T.}~\bibnamefont
  {Maruyama}} \emph {et~al.},\ }\href {\doibase 10.1103/PhysRevC.57.655}
  {\bibfield  {journal} {\bibinfo  {journal} {Phys. Rev. C}\ }\textbf {\bibinfo
  {volume} {57}},\ \bibinfo {pages} {655} (\bibinfo {year} {1998})}\BibitemShut
  {NoStop}%
\bibitem [{\citenamefont {Sonoda}\ \emph {et~al.}(2008)\citenamefont {Sonoda},
  \citenamefont {Watanabe}, \citenamefont {Sato}, \citenamefont {Yasuoka},\
  and\ \citenamefont {Ebisuzaki}}]{Sonoda2008}%
  \BibitemOpen
  \bibfield  {author} {\bibinfo {author} {\bibfnamefont {H.}~\bibnamefont
  {Sonoda}}, \bibinfo {author} {\bibfnamefont {G.}~\bibnamefont {Watanabe}},
  \bibinfo {author} {\bibfnamefont {K.}~\bibnamefont {Sato}}, \bibinfo {author}
  {\bibfnamefont {K.}~\bibnamefont {Yasuoka}}, \ and\ \bibinfo {author}
  {\bibfnamefont {T.}~\bibnamefont {Ebisuzaki}},\ }\href {\doibase
  10.1103/PhysRevC.77.035806} {\bibfield  {journal} {\bibinfo  {journal} {Phys.
  Rev. C}\ }\textbf {\bibinfo {volume} {77}},\ \bibinfo {pages} {035806}
  (\bibinfo {year} {2008})}\BibitemShut {NoStop}%
\bibitem [{\citenamefont {Watanabe}\ \emph {et~al.}(2009)\citenamefont
  {Watanabe}, \citenamefont {Sonoda}, \citenamefont {Maruyama}, \citenamefont
  {Sato}, \citenamefont {Yasuoka},\ and\ \citenamefont
  {Ebisuzaki}}]{Watanabe09}%
  \BibitemOpen
  \bibfield  {author} {\bibinfo {author} {\bibfnamefont {G.}~\bibnamefont
  {Watanabe}}, \bibinfo {author} {\bibfnamefont {H.}~\bibnamefont {Sonoda}},
  \bibinfo {author} {\bibfnamefont {T.}~\bibnamefont {Maruyama}}, \bibinfo
  {author} {\bibfnamefont {K.}~\bibnamefont {Sato}}, \bibinfo {author}
  {\bibfnamefont {K.}~\bibnamefont {Yasuoka}}, \ and\ \bibinfo {author}
  {\bibfnamefont {T.}~\bibnamefont {Ebisuzaki}},\ }\href@noop {} {\bibfield
  {journal} {\bibinfo  {journal} {Phys. Rev. Lett.}\ }\textbf {\bibinfo
  {volume} {103}},\ \bibinfo {pages} {121101} (\bibinfo {year}
  {2009})}\BibitemShut {NoStop}%
\bibitem [{\citenamefont {Bender}\ \emph {et~al.}(2003)\citenamefont {Bender},
  \citenamefont {Heenen},\ and\ \citenamefont {Reinhard}}]{Bender03}%
  \BibitemOpen
  \bibfield  {author} {\bibinfo {author} {\bibfnamefont {M.}~\bibnamefont
  {Bender}}, \bibinfo {author} {\bibfnamefont {P.-H.}\ \bibnamefont {Heenen}},
  \ and\ \bibinfo {author} {\bibfnamefont {P.-G.}\ \bibnamefont {Reinhard}},\
  }\href {\doibase 10.1103/RevModPhys.75.121} {\bibfield  {journal} {\bibinfo
  {journal} {Rev. Mod. Phys.}\ }\textbf {\bibinfo {volume} {75}},\ \bibinfo
  {pages} {121} (\bibinfo {year} {2003})}\BibitemShut {NoStop}%
\bibitem [{\citenamefont {Bonche}\ and\ \citenamefont
  {Vautherin}(1981)}]{Bonche}%
  \BibitemOpen
  \bibfield  {author} {\bibinfo {author} {\bibfnamefont {P.}~\bibnamefont
  {Bonche}}\ and\ \bibinfo {author} {\bibfnamefont {D.}~\bibnamefont
  {Vautherin}},\ }\href {\doibase 10.1016/0375-9474(81)90049-X} {\bibfield
  {journal} {\bibinfo  {journal} {Nucl. Phys. A}\ }\textbf {\bibinfo {volume}
  {372}},\ \bibinfo {pages} {496 } (\bibinfo {year} {1981})}\BibitemShut
  {NoStop}%
\bibitem [{\citenamefont {Magierski}\ and\ \citenamefont
  {Heenen}(2002)}]{Mag02}%
  \BibitemOpen
  \bibfield  {author} {\bibinfo {author} {\bibfnamefont {P.}~\bibnamefont
  {Magierski}}\ and\ \bibinfo {author} {\bibfnamefont {P.-H.}\ \bibnamefont
  {Heenen}},\ }\href {\doibase 10.1103/PhysRevC.65.045804} {\bibfield
  {journal} {\bibinfo  {journal} {Phys. Rev. C}\ }\textbf {\bibinfo {volume}
  {65}},\ \bibinfo {pages} {045804} (\bibinfo {year} {2002})}\BibitemShut
  {NoStop}%
\bibitem [{\citenamefont {G\"ogelein}\ and\ \citenamefont
  {M\"uther}(2007)}]{Goegelein}%
  \BibitemOpen
  \bibfield  {author} {\bibinfo {author} {\bibfnamefont {P.}~\bibnamefont
  {G\"ogelein}}\ and\ \bibinfo {author} {\bibfnamefont {H.}~\bibnamefont
  {M\"uther}},\ }\href {\doibase 10.1103/PhysRevC.76.024312} {\bibfield
  {journal} {\bibinfo  {journal} {Phys. Rev. C}\ }\textbf {\bibinfo {volume}
  {76}},\ \bibinfo {pages} {024312} (\bibinfo {year} {2007})}\BibitemShut
  {NoStop}%
\bibitem [{\citenamefont {Newton}\ and\ \citenamefont
  {Stone}(2009)}]{NewtonStone}%
  \BibitemOpen
  \bibfield  {author} {\bibinfo {author} {\bibfnamefont {W.~G.}\ \bibnamefont
  {Newton}}\ and\ \bibinfo {author} {\bibfnamefont {J.~R.}\ \bibnamefont
  {Stone}},\ }\href@noop {} {\bibfield  {journal} {\bibinfo  {journal} {Phys.
  Rev. C}\ }\textbf {\bibinfo {volume} {79}},\ \bibinfo {pages} {055801}
  (\bibinfo {year} {2009})}\BibitemShut {NoStop}%
\bibitem [{\citenamefont {Pais}\ and\ \citenamefont {Stone}(2012)}]{Pais12}%
  \BibitemOpen
  \bibfield  {author} {\bibinfo {author} {\bibfnamefont {H.}~\bibnamefont
  {Pais}}\ and\ \bibinfo {author} {\bibfnamefont {J.~R.}\ \bibnamefont
  {Stone}},\ }\href {\doibase 10.1103/PhysRevLett.109.151101} {\bibfield
  {journal} {\bibinfo  {journal} {Phys. Rev. Lett.}\ }\textbf {\bibinfo
  {volume} {109}},\ \bibinfo {pages} {151101} (\bibinfo {year}
  {2012})}\BibitemShut {NoStop}%
\bibitem [{\citenamefont {Schuetrumpf}\ \emph {et~al.}(2013)\citenamefont
  {Schuetrumpf}, \citenamefont {Klatt}, \citenamefont {Iida}, \citenamefont
  {Maruhn}, \citenamefont {Mecke},\ and\ \citenamefont
  {Reinhard}}]{Schuetrumpf2013a}%
  \BibitemOpen
  \bibfield  {author} {\bibinfo {author} {\bibfnamefont {B.}~\bibnamefont
  {Schuetrumpf}}, \bibinfo {author} {\bibfnamefont {M.~A.}\ \bibnamefont
  {Klatt}}, \bibinfo {author} {\bibfnamefont {K.}~\bibnamefont {Iida}},
  \bibinfo {author} {\bibfnamefont {J.~A.}\ \bibnamefont {Maruhn}}, \bibinfo
  {author} {\bibfnamefont {K.}~\bibnamefont {Mecke}}, \ and\ \bibinfo {author}
  {\bibfnamefont {P.-G.}\ \bibnamefont {Reinhard}},\ }\href {\doibase
  10.1103/PhysRevC.87.055805} {\bibfield  {journal} {\bibinfo  {journal} {Phys.
  Rev. C}\ }\textbf {\bibinfo {volume} {87}},\ \bibinfo {pages} {055805}
  (\bibinfo {year} {2013})}\BibitemShut {NoStop}%
\bibitem [{\citenamefont {Schuetrumpf}\ \emph {et~al.}(2014)\citenamefont
  {Schuetrumpf}, \citenamefont {Iida}, \citenamefont {Maruhn},\ and\
  \citenamefont {Reinhard}}]{Schuetrumpf2014}%
  \BibitemOpen
  \bibfield  {author} {\bibinfo {author} {\bibfnamefont {B.}~\bibnamefont
  {Schuetrumpf}}, \bibinfo {author} {\bibfnamefont {K.}~\bibnamefont {Iida}},
  \bibinfo {author} {\bibfnamefont {J.~A.}\ \bibnamefont {Maruhn}}, \ and\
  \bibinfo {author} {\bibfnamefont {P.-G.}\ \bibnamefont {Reinhard}},\ }\href
  {\doibase 10.1103/PhysRevC.90.055802} {\bibfield  {journal} {\bibinfo
  {journal} {Phys. Rev. C}\ }\textbf {\bibinfo {volume} {90}},\ \bibinfo
  {pages} {055802} (\bibinfo {year} {2014})}\BibitemShut {NoStop}%
\bibitem [{\citenamefont {Schuetrumpf}\ \emph {et~al.}(2015)\citenamefont
  {Schuetrumpf}, \citenamefont {Klatt}, \citenamefont {Iida}, \citenamefont
  {Schr\"oder-Turk}, \citenamefont {Maruhn}, \citenamefont {Mecke},\ and\
  \citenamefont {Reinhard}}]{Schuetrumpf2015}%
  \BibitemOpen
  \bibfield  {author} {\bibinfo {author} {\bibfnamefont {B.}~\bibnamefont
  {Schuetrumpf}}, \bibinfo {author} {\bibfnamefont {M.~A.}\ \bibnamefont
  {Klatt}}, \bibinfo {author} {\bibfnamefont {K.}~\bibnamefont {Iida}},
  \bibinfo {author} {\bibfnamefont {G.~E.}\ \bibnamefont {Schr\"oder-Turk}},
  \bibinfo {author} {\bibfnamefont {J.~A.}\ \bibnamefont {Maruhn}}, \bibinfo
  {author} {\bibfnamefont {K.}~\bibnamefont {Mecke}}, \ and\ \bibinfo {author}
  {\bibfnamefont {P.-G.}\ \bibnamefont {Reinhard}},\ }\href {\doibase
  10.1103/PhysRevC.91.025801} {\bibfield  {journal} {\bibinfo  {journal} {Phys.
  Rev. C}\ }\textbf {\bibinfo {volume} {91}},\ \bibinfo {pages} {025801}
  (\bibinfo {year} {2015})}\BibitemShut {NoStop}%
\bibitem [{\citenamefont {Horowitz}\ \emph {et~al.}(2015)\citenamefont
  {Horowitz}, \citenamefont {Berry}, \citenamefont {Briggs}, \citenamefont
  {Caplan}, \citenamefont {Cumming},\ and\ \citenamefont
  {Schneider}}]{Horowitz2015}%
  \BibitemOpen
  \bibfield  {author} {\bibinfo {author} {\bibfnamefont {C.~J.}\ \bibnamefont
  {Horowitz}}, \bibinfo {author} {\bibfnamefont {D.~K.}\ \bibnamefont {Berry}},
  \bibinfo {author} {\bibfnamefont {C.~M.}\ \bibnamefont {Briggs}}, \bibinfo
  {author} {\bibfnamefont {M.~E.}\ \bibnamefont {Caplan}}, \bibinfo {author}
  {\bibfnamefont {A.}~\bibnamefont {Cumming}}, \ and\ \bibinfo {author}
  {\bibfnamefont {A.~S.}\ \bibnamefont {Schneider}},\ }\href {\doibase
  10.1103/PhysRevLett.114.031102} {\bibfield  {journal} {\bibinfo  {journal}
  {Phys. Rev. Lett.}\ }\textbf {\bibinfo {volume} {114}},\ \bibinfo {pages}
  {031102} (\bibinfo {year} {2015})}\BibitemShut {NoStop}%
\bibitem [{\citenamefont {Gros}(1992)}]{Gros1992}%
  \BibitemOpen
  \bibfield  {author} {\bibinfo {author} {\bibfnamefont {C.}~\bibnamefont
  {Gros}},\ }\href@noop {} {\bibfield  {journal} {\bibinfo  {journal} {Z. Phys.
  B}\ }\textbf {\bibinfo {volume} {86}},\ \bibinfo {pages} {359} (\bibinfo
  {year} {1992})}\BibitemShut {NoStop}%
\bibitem [{\citenamefont {Gros}(1996)}]{Gros1996}%
  \BibitemOpen
  \bibfield  {author} {\bibinfo {author} {\bibfnamefont {C.}~\bibnamefont
  {Gros}},\ }\href {\doibase 10.1103/PhysRevB.53.6865} {\bibfield  {journal}
  {\bibinfo  {journal} {Phys. Rev. B}\ }\textbf {\bibinfo {volume} {53}},\
  \bibinfo {pages} {6865} (\bibinfo {year} {1996})}\BibitemShut {NoStop}%
\bibitem [{\citenamefont {Lin}\ \emph {et~al.}(2001)\citenamefont {Lin},
  \citenamefont {Zong},\ and\ \citenamefont {Ceperley}}]{Lin}%
  \BibitemOpen
  \bibfield  {author} {\bibinfo {author} {\bibfnamefont {C.}~\bibnamefont
  {Lin}}, \bibinfo {author} {\bibfnamefont {F.~H.}\ \bibnamefont {Zong}}, \
  and\ \bibinfo {author} {\bibfnamefont {D.~M.}\ \bibnamefont {Ceperley}},\
  }\href {\doibase 10.1103/PhysRevE.64.016702} {\bibfield  {journal} {\bibinfo
  {journal} {Phys. Rev. E}\ }\textbf {\bibinfo {volume} {64}},\ \bibinfo
  {pages} {016702} (\bibinfo {year} {2001})}\BibitemShut {NoStop}%
\bibitem [{\citenamefont {Hagen}\ \emph {et~al.}(2014)\citenamefont {Hagen},
  \citenamefont {Papenbrock}, \citenamefont {Ekstr{\"o}m}, \citenamefont
  {Wendt}, \citenamefont {Baardsen}, \citenamefont {Gandolfi}, \citenamefont
  {Hjorth-Jensen},\ and\ \citenamefont {Horowitz}}]{Hagen14}%
  \BibitemOpen
  \bibfield  {author} {\bibinfo {author} {\bibfnamefont {G.}~\bibnamefont
  {Hagen}}, \bibinfo {author} {\bibfnamefont {T.}~\bibnamefont {Papenbrock}},
  \bibinfo {author} {\bibfnamefont {A.}~\bibnamefont {Ekstr{\"o}m}}, \bibinfo
  {author} {\bibfnamefont {K.~A.}\ \bibnamefont {Wendt}}, \bibinfo {author}
  {\bibfnamefont {G.}~\bibnamefont {Baardsen}}, \bibinfo {author}
  {\bibfnamefont {S.}~\bibnamefont {Gandolfi}}, \bibinfo {author}
  {\bibfnamefont {M.}~\bibnamefont {Hjorth-Jensen}}, \ and\ \bibinfo {author}
  {\bibfnamefont {C.~J.}\ \bibnamefont {Horowitz}},\ }\href {\doibase
  10.1103/PhysRevC.89.014319} {\bibfield  {journal} {\bibinfo  {journal} {Phys.
  Rev. C}\ }\textbf {\bibinfo {volume} {89}},\ \bibinfo {pages} {014319}
  (\bibinfo {year} {2014})}\BibitemShut {NoStop}%
\bibitem [{\citenamefont {Tiburzi}(2005)}]{Tiburzi05}%
  \BibitemOpen
  \bibfield  {author} {\bibinfo {author} {\bibfnamefont {B.~C.}\ \bibnamefont
  {Tiburzi}},\ }\href
  {http://www.sciencedirect.com/science/article/pii/S0370269305006118}
  {\bibfield  {journal} {\bibinfo  {journal} {Phys. Lett. B}\ }\textbf
  {\bibinfo {volume} {617}},\ \bibinfo {pages} {40 } (\bibinfo {year}
  {2005})}\BibitemShut {NoStop}%
\bibitem [{\citenamefont {Brice{\~n}o}\ \emph {et~al.}(2014)\citenamefont
  {Brice{\~n}o}, \citenamefont {Davoudi}, \citenamefont {Luu},\ and\
  \citenamefont {Savage}}]{Briceno14}%
  \BibitemOpen
  \bibfield  {author} {\bibinfo {author} {\bibfnamefont {R.~A.}\ \bibnamefont
  {Brice{\~n}o}}, \bibinfo {author} {\bibfnamefont {Z.}~\bibnamefont
  {Davoudi}}, \bibinfo {author} {\bibfnamefont {T.~C.}\ \bibnamefont {Luu}}, \
  and\ \bibinfo {author} {\bibfnamefont {M.~J.}\ \bibnamefont {Savage}},\
  }\href {\doibase 10.1103/PhysRevD.89.074509} {\bibfield  {journal} {\bibinfo
  {journal} {Phys. Rev. D}\ }\textbf {\bibinfo {volume} {89}},\ \bibinfo
  {pages} {074509} (\bibinfo {year} {2014})}\BibitemShut {NoStop}%
\bibitem [{\citenamefont {Carter}\ \emph {et~al.}(2005)\citenamefont {Carter},
  \citenamefont {Chamel},\ and\ \citenamefont {Haensel}}]{Carter}%
  \BibitemOpen
  \bibfield  {author} {\bibinfo {author} {\bibfnamefont {B.}~\bibnamefont
  {Carter}}, \bibinfo {author} {\bibfnamefont {N.}~\bibnamefont {Chamel}}, \
  and\ \bibinfo {author} {\bibfnamefont {P.}~\bibnamefont {Haensel}},\ }\href
  {\doibase http://dx.doi.org/10.1016/j.nuclphysa.2004.11.006} {\bibfield
  {journal} {\bibinfo  {journal} {Nuclear Physics A}\ }\textbf {\bibinfo
  {volume} {748}},\ \bibinfo {pages} {675 } (\bibinfo {year}
  {2005})}\BibitemShut {NoStop}%
\bibitem [{\citenamefont {Chamel}\ \emph {et~al.}(2007)\citenamefont {Chamel},
  \citenamefont {Naimi}, \citenamefont {Khan},\ and\ \citenamefont
  {Margueron}}]{Chamel2007}%
  \BibitemOpen
  \bibfield  {author} {\bibinfo {author} {\bibfnamefont {N.}~\bibnamefont
  {Chamel}}, \bibinfo {author} {\bibfnamefont {S.}~\bibnamefont {Naimi}},
  \bibinfo {author} {\bibfnamefont {E.}~\bibnamefont {Khan}}, \ and\ \bibinfo
  {author} {\bibfnamefont {J.}~\bibnamefont {Margueron}},\ }\href {\doibase
  10.1103/PhysRevC.75.055806} {\bibfield  {journal} {\bibinfo  {journal} {Phys.
  Rev. C}\ }\textbf {\bibinfo {volume} {75}},\ \bibinfo {pages} {055806}
  (\bibinfo {year} {2007})}\BibitemShut {NoStop}%
\bibitem [{\citenamefont {Gulminelli}\ \emph {et~al.}(2011)\citenamefont
  {Gulminelli}, \citenamefont {Furuta}, \citenamefont {Juillet},\ and\
  \citenamefont {Leclercq}}]{Gulminelli}%
  \BibitemOpen
  \bibfield  {author} {\bibinfo {author} {\bibfnamefont {F.}~\bibnamefont
  {Gulminelli}}, \bibinfo {author} {\bibfnamefont {T.}~\bibnamefont {Furuta}},
  \bibinfo {author} {\bibfnamefont {O.}~\bibnamefont {Juillet}}, \ and\
  \bibinfo {author} {\bibfnamefont {C.}~\bibnamefont {Leclercq}},\ }\href
  {\doibase 10.1103/PhysRevC.84.065806} {\bibfield  {journal} {\bibinfo
  {journal} {Phys. Rev. C}\ }\textbf {\bibinfo {volume} {84}},\ \bibinfo
  {pages} {065806} (\bibinfo {year} {2011})}\BibitemShut {NoStop}%
\bibitem [{\citenamefont {Maruhn}\ \emph {et~al.}(2014)\citenamefont {Maruhn},
  \citenamefont {Reinhard}, \citenamefont {Stevenson},\ and\ \citenamefont
  {Umar}}]{Mar15a}%
  \BibitemOpen
  \bibfield  {author} {\bibinfo {author} {\bibfnamefont {J.}~\bibnamefont
  {Maruhn}}, \bibinfo {author} {\bibfnamefont {P.-G.}\ \bibnamefont
  {Reinhard}}, \bibinfo {author} {\bibfnamefont {P.}~\bibnamefont {Stevenson}},
  \ and\ \bibinfo {author} {\bibfnamefont {A.}~\bibnamefont {Umar}},\ }\href
  {\doibase http://dx.doi.org/10.1016/j.cpc.2014.04.008} {\bibfield  {journal}
  {\bibinfo  {journal} {Comput. Phys. Commun.}\ }\textbf {\bibinfo {volume}
  {185}},\ \bibinfo {pages} {2195 } (\bibinfo {year} {2014})}\BibitemShut
  {NoStop}%
\bibitem [{\citenamefont {Chabanat}\ \emph {et~al.}(1998)\citenamefont
  {Chabanat} \emph {et~al.}}]{Chabanat}%
  \BibitemOpen
  \bibfield  {author} {\bibinfo {author} {\bibfnamefont {E.}~\bibnamefont
  {Chabanat}} \emph {et~al.},\ }\href@noop {} {\bibfield  {journal} {\bibinfo
  {journal} {Nucl. Phys. A}\ }\textbf {\bibinfo {volume} {635}},\ \bibinfo
  {pages} {231 } (\bibinfo {year} {1998})}\BibitemShut {NoStop}%
\bibitem [{\citenamefont {Alcain}\ \emph {et~al.}(2014)\citenamefont {Alcain},
  \citenamefont {Gim\'enez~Molinelli}, \citenamefont {Nichols},\ and\
  \citenamefont {Dorso}}]{Alcain}%
  \BibitemOpen
  \bibfield  {author} {\bibinfo {author} {\bibfnamefont {P.~N.}\ \bibnamefont
  {Alcain}}, \bibinfo {author} {\bibfnamefont {P.~A.}\ \bibnamefont
  {Gim\'enez~Molinelli}}, \bibinfo {author} {\bibfnamefont {J.~I.}\
  \bibnamefont {Nichols}}, \ and\ \bibinfo {author} {\bibfnamefont {C.~O.}\
  \bibnamefont {Dorso}},\ }\href {\doibase 10.1103/PhysRevC.89.055801}
  {\bibfield  {journal} {\bibinfo  {journal} {Phys. Rev. C}\ }\textbf {\bibinfo
  {volume} {89}},\ \bibinfo {pages} {055801} (\bibinfo {year}
  {2014})}\BibitemShut {NoStop}%
\bibitem [{\citenamefont {Schnering}\ and\ \citenamefont
  {Nesper}(1991)}]{NesperSchnering1991}%
  \BibitemOpen
  \bibfield  {author} {\bibinfo {author} {\bibfnamefont {H.}~\bibnamefont
  {Schnering}}\ and\ \bibinfo {author} {\bibfnamefont {R.}~\bibnamefont
  {Nesper}},\ }\href {\doibase 10.1007/BF01313411} {\bibfield  {journal}
  {\bibinfo  {journal} {Z. Phys. B}\ }\textbf {\bibinfo {volume} {83}},\
  \bibinfo {pages} {407} (\bibinfo {year} {1991})}\BibitemShut {NoStop}%
\bibitem [{\citenamefont {Nakazato}\ \emph {et~al.}(2009)\citenamefont
  {Nakazato}, \citenamefont {Oyamatsu},\ and\ \citenamefont
  {Yamada}}]{Nakazato2009}%
  \BibitemOpen
  \bibfield  {author} {\bibinfo {author} {\bibfnamefont {K.}~\bibnamefont
  {Nakazato}}, \bibinfo {author} {\bibfnamefont {K.}~\bibnamefont {Oyamatsu}},
  \ and\ \bibinfo {author} {\bibfnamefont {S.}~\bibnamefont {Yamada}},\ }\href
  {\doibase 10.1103/PhysRevLett.103.132501} {\bibfield  {journal} {\bibinfo
  {journal} {Phys. Rev. Lett.}\ }\textbf {\bibinfo {volume} {103}},\ \bibinfo
  {pages} {132501} (\bibinfo {year} {2009})}\BibitemShut {NoStop}%
\bibitem [{\citenamefont {Nakazato}\ \emph {et~al.}(2011)\citenamefont
  {Nakazato}, \citenamefont {Iida},\ and\ \citenamefont
  {Oyamatsu}}]{Nakazato2011}%
  \BibitemOpen
  \bibfield  {author} {\bibinfo {author} {\bibfnamefont {K.}~\bibnamefont
  {Nakazato}}, \bibinfo {author} {\bibfnamefont {K.}~\bibnamefont {Iida}}, \
  and\ \bibinfo {author} {\bibfnamefont {K.}~\bibnamefont {Oyamatsu}},\ }\href
  {\doibase 10.1103/PhysRevC.83.065811} {\bibfield  {journal} {\bibinfo
  {journal} {Phys. Rev. C}\ }\textbf {\bibinfo {volume} {83}},\ \bibinfo
  {pages} {065811} (\bibinfo {year} {2011})}\BibitemShut {NoStop}%
\bibitem [{\citenamefont {Michielsen}\ and\ \citenamefont
  {Stavenga}(2008)}]{MichielsenStavenga:2008}%
  \BibitemOpen
  \bibfield  {author} {\bibinfo {author} {\bibfnamefont {K.}~\bibnamefont
  {Michielsen}}\ and\ \bibinfo {author} {\bibfnamefont {D.}~\bibnamefont
  {Stavenga}},\ }\href@noop {} {\bibfield  {journal} {\bibinfo  {journal} {J.
  R. Soc. Interface}\ }\textbf {\bibinfo {volume} {5}},\ \bibinfo {pages} {85}
  (\bibinfo {year} {2008})}\BibitemShut {NoStop}%
\bibitem [{\citenamefont {Saranathan}\ \emph {et~al.}(2010)\citenamefont
  {Saranathan}, \citenamefont {Osuji}, \citenamefont {Mochrie}, \citenamefont
  {Noh}, \citenamefont {Narayanan}, \citenamefont {Sandy}, \citenamefont
  {Dufresne},\ and\ \citenamefont
  {Prum}}]{SaranathanOsujiMochrieNohNarayananSandyDufresnePrum:2010}%
  \BibitemOpen
  \bibfield  {author} {\bibinfo {author} {\bibfnamefont {V.}~\bibnamefont
  {Saranathan}}, \bibinfo {author} {\bibfnamefont {C.}~\bibnamefont {Osuji}},
  \bibinfo {author} {\bibfnamefont {S.}~\bibnamefont {Mochrie}}, \bibinfo
  {author} {\bibfnamefont {H.}~\bibnamefont {Noh}}, \bibinfo {author}
  {\bibfnamefont {S.}~\bibnamefont {Narayanan}}, \bibinfo {author}
  {\bibfnamefont {A.}~\bibnamefont {Sandy}}, \bibinfo {author} {\bibfnamefont
  {E.}~\bibnamefont {Dufresne}}, \ and\ \bibinfo {author} {\bibfnamefont
  {R.}~\bibnamefont {Prum}},\ }\href {\doibase 10.1073/pnas.0909616107}
  {\bibfield  {journal} {\bibinfo  {journal} {PNAS}\ }\textbf {\bibinfo
  {volume} {107}},\ \bibinfo {pages} {11676} (\bibinfo {year}
  {2010})}\BibitemShut {NoStop}%
\bibitem [{\citenamefont {Schr{\"o}der-Turk}\ \emph {et~al.}(2011)\citenamefont
  {Schr{\"o}der-Turk}, \citenamefont {Wickham}, \citenamefont {Averdunk},
  \citenamefont {Large}, \citenamefont {Poladian}, \citenamefont {Brink},
  \citenamefont {{Fitz Gerald}},\ and\ \citenamefont
  {Hyde}}]{SchroederTurkWickhamAverdunkBrinkFitzGeraldPoladianLargeHyde:2011}%
  \BibitemOpen
  \bibfield  {author} {\bibinfo {author} {\bibfnamefont {G.}~\bibnamefont
  {Schr{\"o}der-Turk}}, \bibinfo {author} {\bibfnamefont {S.}~\bibnamefont
  {Wickham}}, \bibinfo {author} {\bibfnamefont {H.}~\bibnamefont {Averdunk}},
  \bibinfo {author} {\bibfnamefont {M.}~\bibnamefont {Large}}, \bibinfo
  {author} {\bibfnamefont {L.}~\bibnamefont {Poladian}}, \bibinfo {author}
  {\bibfnamefont {F.}~\bibnamefont {Brink}}, \bibinfo {author} {\bibfnamefont
  {J.}~\bibnamefont {{Fitz Gerald}}}, \ and\ \bibinfo {author} {\bibfnamefont
  {S.~T.}\ \bibnamefont {Hyde}},\ }\href {\doibase
  doi:10.1016/j.jsb.2011.01.004} {\bibfield  {journal} {\bibinfo  {journal}
  {J.~Struct.~Biol.}\ }\textbf {\bibinfo {volume} {174}},\ \bibinfo {pages}
  {290} (\bibinfo {year} {2011})}\BibitemShut {NoStop}%
\bibitem [{\citenamefont {Hajduk}\ \emph {et~al.}(1994)\citenamefont {Hajduk},
  \citenamefont {Harper}, \citenamefont {Gruner}, \citenamefont {Honeker},
  \citenamefont {Kim}, \citenamefont {Thomas},\ and\ \citenamefont
  {Fetters}}]{Hajduk1994}%
  \BibitemOpen
  \bibfield  {author} {\bibinfo {author} {\bibfnamefont {D.}~\bibnamefont
  {Hajduk}}, \bibinfo {author} {\bibfnamefont {P.}~\bibnamefont {Harper}},
  \bibinfo {author} {\bibfnamefont {S.}~\bibnamefont {Gruner}}, \bibinfo
  {author} {\bibfnamefont {C.}~\bibnamefont {Honeker}}, \bibinfo {author}
  {\bibfnamefont {G.}~\bibnamefont {Kim}}, \bibinfo {author} {\bibfnamefont
  {E.}~\bibnamefont {Thomas}}, \ and\ \bibinfo {author} {\bibfnamefont
  {L.}~\bibnamefont {Fetters}},\ }\href@noop {} {\bibfield  {journal} {\bibinfo
   {journal} {Macromolecules}\ }\textbf {\bibinfo {volume} {27}},\ \bibinfo
  {pages} {4063} (\bibinfo {year} {1994})}\BibitemShut {NoStop}%
\bibitem [{\citenamefont {Larsson}(1989)}]{Larsson:1989}%
  \BibitemOpen
  \bibfield  {author} {\bibinfo {author} {\bibfnamefont {K.}~\bibnamefont
  {Larsson}},\ }\href@noop {} {\bibfield  {journal} {\bibinfo  {journal} {J.
  Phys. Chem.}\ }\textbf {\bibinfo {volume} {93}},\ \bibinfo {pages} {7304}
  (\bibinfo {year} {1989})}\BibitemShut {NoStop}%
\bibitem [{\citenamefont {Hyde}\ \emph {et~al.}(1997)\citenamefont {Hyde},
  \citenamefont {Andersson}, \citenamefont {Larsson}, \citenamefont {Blum},
  \citenamefont {Landh}, \citenamefont {Lidin},\ and\ \citenamefont
  {Ninham}}]{HydeLanguageOfShape:1997}%
  \BibitemOpen
  \bibfield  {author} {\bibinfo {author} {\bibfnamefont {S.~T.}\ \bibnamefont
  {Hyde}}, \bibinfo {author} {\bibfnamefont {S.}~\bibnamefont {Andersson}},
  \bibinfo {author} {\bibfnamefont {K.}~\bibnamefont {Larsson}}, \bibinfo
  {author} {\bibfnamefont {Z.}~\bibnamefont {Blum}}, \bibinfo {author}
  {\bibfnamefont {T.}~\bibnamefont {Landh}}, \bibinfo {author} {\bibfnamefont
  {S.}~\bibnamefont {Lidin}}, \ and\ \bibinfo {author} {\bibfnamefont
  {B.}~\bibnamefont {Ninham}},\ }\href@noop {} {\emph {\bibinfo {title} {The
  Language of Shape}}},\ \bibinfo {edition} {1st}\ ed.\ (\bibinfo  {publisher}
  {Elsevier Science},\ \bibinfo {address} {Amsterdam},\ \bibinfo {year}
  {1997})\BibitemShut {NoStop}%
\bibitem [{\citenamefont {Schr\"oder}\ \emph {et~al.}(2003)\citenamefont
  {Schr\"oder}, \citenamefont {Ramsden}, \citenamefont {Christy},\ and\
  \citenamefont {Hyde}}]{Schroeder2003}%
  \BibitemOpen
  \bibfield  {author} {\bibinfo {author} {\bibfnamefont {G.~E.}\ \bibnamefont
  {Schr\"oder}}, \bibinfo {author} {\bibfnamefont {S.~J.}\ \bibnamefont
  {Ramsden}}, \bibinfo {author} {\bibfnamefont {A.~G.}\ \bibnamefont
  {Christy}}, \ and\ \bibinfo {author} {\bibfnamefont {S.~T.}\ \bibnamefont
  {Hyde}},\ }\href@noop {} {\bibfield  {journal} {\bibinfo  {journal} {Eur.
  Phys. J. B}\ }\textbf {\bibinfo {volume} {35}},\ \bibinfo {pages} {551}
  (\bibinfo {year} {2003})}\BibitemShut {NoStop}%
\bibitem [{\citenamefont {Fogden}\ and\ \citenamefont
  {Hyde}(1999)}]{FogdenHyde:1999}%
  \BibitemOpen
  \bibfield  {author} {\bibinfo {author} {\bibfnamefont {A.}~\bibnamefont
  {Fogden}}\ and\ \bibinfo {author} {\bibfnamefont {S.~T.}\ \bibnamefont
  {Hyde}},\ }\href@noop {} {\bibfield  {journal} {\bibinfo  {journal} {Eur.
  Phys. J. B}\ }\textbf {\bibinfo {volume} {7}},\ \bibinfo {pages} {91}
  (\bibinfo {year} {1999})}\BibitemShut {NoStop}%
\bibitem [{\citenamefont {Schr{\"o}der-Turk}\ \emph {et~al.}(2006)\citenamefont
  {Schr{\"o}der-Turk}, \citenamefont {Fogden},\ and\ \citenamefont
  {Hyde}}]{SchroederFogdenHyde:2006}%
  \BibitemOpen
  \bibfield  {author} {\bibinfo {author} {\bibfnamefont {G.~E.}\ \bibnamefont
  {Schr{\"o}der-Turk}}, \bibinfo {author} {\bibfnamefont {A.}~\bibnamefont
  {Fogden}}, \ and\ \bibinfo {author} {\bibfnamefont {S.~T.}\ \bibnamefont
  {Hyde}},\ }\href@noop {} {\bibfield  {journal} {\bibinfo  {journal} {Eur.
  Phys. J B}\ }\textbf {\bibinfo {volume} {54}},\ \bibinfo {pages} {509}
  (\bibinfo {year} {2006})}\BibitemShut {NoStop}%
\bibitem [{\citenamefont {Pei}\ \emph {et~al.}(2014)\citenamefont {Pei},
  \citenamefont {Fann}, \citenamefont {Harrison}, \citenamefont {Nazarewicz},
  \citenamefont {Shi},\ and\ \citenamefont {Thornton}}]{Pei2014}%
  \BibitemOpen
  \bibfield  {author} {\bibinfo {author} {\bibfnamefont {J.~C.}\ \bibnamefont
  {Pei}}, \bibinfo {author} {\bibfnamefont {G.~I.}\ \bibnamefont {Fann}},
  \bibinfo {author} {\bibfnamefont {R.~J.}\ \bibnamefont {Harrison}}, \bibinfo
  {author} {\bibfnamefont {W.}~\bibnamefont {Nazarewicz}}, \bibinfo {author}
  {\bibfnamefont {Y.}~\bibnamefont {Shi}}, \ and\ \bibinfo {author}
  {\bibfnamefont {S.}~\bibnamefont {Thornton}},\ }\href {\doibase
  10.1103/PhysRevC.90.024317} {\bibfield  {journal} {\bibinfo  {journal} {Phys.
  Rev. C}\ }\textbf {\bibinfo {volume} {90}},\ \bibinfo {pages} {024317}
  (\bibinfo {year} {2014})}\BibitemShut {NoStop}%
\end{thebibliography}%

\end{document}